\documentstyle[12pt]{article}
\begin{document}
\input{epsf}
\def\epp{\epsilon^{\prime}}
\def\vep{\varepsilon}
\def\ra{\rightarrow}
\def\ppg{\pi^+\pi^-\gamma}
\def\vp{{\bf p}}
\def\ko{K^0}
\def\kb{\bar{K^0}}
\def\al{\alpha}
\def\ab{\bar{\alpha}}
\def\be{\begin{equation}}
\def\ee{\end{equation}}
\def\bea{\begin{eqnarray}}
\def\eea{\end{eqnarray}}
\def\question#1{{{\marginpar{\small \sc #1}}}}
\def\mpl{{{m_{Pl}}}}
\def\oph{{{\Omega_{\widetilde{\gamma}}h^2}}}
\def\be{\begin{equation}}
\def\ee{\end{equation}}
\def\ba{\begin{eqnarray}}
\def\ea{\end{eqnarray}}
\def\la{\mathrel{\mathpalette\fun <}}
\def\ga{\mathrel{\mathpalette\fun >}}
\def\fun#1#2{\lower3.6pt\vbox{\baselineskip0pt\lineskip.9pt
        \ialign{$\mathsurround=0pt#1\hfill##\hfil$\crcr#2\crcr\sim\crcr}}}
\def\pho{{{\widetilde{\gamma}}}}
\def\r0{{{R^0}}}
\def\glu{{{\widetilde{g}}}}
\def\sec{{{\mbox{sec}}}}
\def\GeV{{{\mbox{GeV}}}}
\def\MeV{{{\mbox{MeV}}}}
\def\SUSY{{{{\sc susy}}}}
\def\LSP{{{{\sc lsp}}}}
\def\LEP{{{{\sc lep}}}}
\def\LROCS{{{{\sc lrocs}}}}
\def\WIMPS{{{{\sc wimps}}}}
\def\cm{{{\mbox{cm}}}}
\def\photino{{{\mbox{photino}}}}
\def\gluino{{{\mbox{gluino}}}}
\def\mb{{{\mbox{mb}}}}
\def\avg#1{{{{\langle #1 \rangle }}}}
\def\taun{{{\tau_{9}}}}
\def\mpl{{{m_{Pl}}}}
\def\re#1{{[\ref{#1}]}}
\def\eqr#1{{Eq.\ (\ref{#1})}}
\def\mst{{{M_{\widetilde{S}}}}}
\newcommand{\gl}{\tilde{g}}
\newcommand{\sneu}{\tilde{\nu}}
\newcommand{\sq}{\tilde{q}}
\newcommand{\se}{\tilde{e}}
\newcommand{\ch}{\chi^{\pm}}
\newcommand{\neut}{\chi^{0}}
\newcommand{\gsi}{\,\raisebox{-0.13cm}{$\stackrel{\textstyle>}
{\textstyle\sim}$}\,}
\newcommand{\lsi}{\,\raisebox{-0.13cm}{$\stackrel{\textstyle<}
{\textstyle\sim}$}\,}

\rightline{RU-97-82}  
\baselineskip=18pt
\vskip 0.7in
\begin{center}
{\bf \LARGE Experimental and Cosmological Implications of Light Gauginos}\\
\vspace*{0.9in}
{\large Glennys R. Farrar}\footnote{Based on talks presented at the 9th Intl.
Baksan School "Particles and Cosmology", INR Baksan, Russia, March 1997
and at the Intl. Workshop on Physics Beyond the Desert, Tegernsee, Germany,
June 1997.  Research supported in part by NSF-PHY-94-2302.}\\ 
{\it Department of Physics and Astronomy}\\
{\it Rutgers University, Piscataway, NJ 08855, USA}\\
\vspace*{0.1in} 
\end{center}

\vskip  0.5in

{\bf Abstract:}
Gauginos may be nearly massless at tree level, with loop corrections giving
a gluino mass of order 100 MeV and a photino mass of order 1 GeV.  Relic
photinos can naturally account for the observed dark matter, but their detection
is more difficult than for conventional WIMPs.  The lightest gluino-containing
baryon could account for the recently observed ultra-high energy cosmic rays,
which violate the GZK bound. The predicted mass and properties of the $\gl
\gl$ boundstate agree with those of the $\eta(1410)$, a flavor singlet
pseudoscalar meson which has proved difficult to reconcile with QCD
predictions.  Laboratory experiments presently exclude only a small portion
of the interesting parameter space, although improvements expected in
the next year may lead to complete exclusion or discovery.

\thispagestyle{empty}
\newpage
\addtocounter{page}{-1}

\section*{Introduction}
\hspace*{2em}
Some supersymmetry (SUSY) breaking scenarios produce negligible
tree-level gaugino masses and scalar trilinear couplings
($m_1=m_2=m_3=A=0$), and conserve $R$-parity.  Such SUSY breaking
has several attractive theoretical consequences such as the
absence of the ``SUSY CP problem"\cite{f99101}.  Although
massless at tree level, gauginos get calculable masses through
radiative corrections from electroweak (gaugino/higgsino-Higgs/gauge
boson) and top-stop loops\cite{radmass}.  Evaluating these within the
constrained parameter space leads to a gluino mass range $m_{\tilde{g}}\sim
\frac{1}{10} - \frac{1}{2}$ GeV\cite{f99101}, while analysis of the
$\eta'$ mass narrow this to $m(\glu) \approx 120$ MeV\cite{f:108}. The
photino mass range depends on more unknowns than the gluino mass, such
as the higgs and higgsino sectors, but can be estimated to be
$m_{\tilde{\gamma}} \sim \frac{1}{10} - 1 \frac{1}{2}$
GeV\cite{f99101}.   

The gluino binds with quarks, antiquarks and/or gluons to make
color-singlet hadrons (generically called $R$-hadrons\cite{f:24}). 
The lightest of these is expected to be the gluino-gluon bound state,
designated $\r0$.  It is predicted to have a mass in the range
$1.3-2.2$ GeV, approximately degenerate with the
lightest glueball ($0^{++}$) and ``gluinoball'' ($0^{-+},~\glu
\glu$)\cite{f:95,f:104}.  An encouraging development for this scenario
is the existance of an ``extra'' isosinglet pseudoscalar
meson, $\eta(1410)$, which is difficult to accomodate in standard QCD
but which matches nicely the properties of the pseudoscalar $\glu
\glu$\cite{f:109,f:104}.  

Due to the non-negligible mass of the photino compared to the $\r0$,
the $\r0$ is long lived.  Its lifetime is estimated to be in the range 
$10^{-10} - 10^{-5}$ sec\cite{f99101}.  Prompt
photinos\cite{f:24} are not a useful signature for the light gluinos
and the energy they carry\cite{f:51}. Thus gluino masses less than
about $ \frac{1}{2}$ GeV are largely
unconstrained\cite{f:95}\footnote{The ALPEH claim to exclude 
light gluinos\cite{aleph:lg} assigns a $1 \sigma$ theoretical
systematic error based on varying the renormalization scale over a
small range. Taking a more generally accepted range of scale variation
and accounting for the large sensitivity to hadronization model, the
ALEPH systematic uncertainty is comparable to that of other
experiments and does not exclude light gluinos\cite{fLaT,f:119}.  The claim
of Nagy and Troscsanyi\cite{nt:lg}, that use of $R_4$ allows a 95\% cl exclusion,
has even worse problems.  In addition to scale sensitivity, their result relies
on using the central value of $\alpha_s$.  When the error bars on $\alpha_s$
are included, their limit is reduced to $1 \sigma$, even without
considering the uncertainty due to scale and resummation scheme sensitivity.}.
Proposals for direct searches for hadrons containing gluinos, via
their decays in $K^0$ beams and otherwise, are given in Refs.
\cite{f:95,f:104}.  Results of two new experiments are mentioned below.
For a recent detailed survey of the experimental constraints on light
gaugino scenarios, see \cite{f:119}.

\section*{Relic Photinos}
\hspace*{2em}
In the light gaugino scenario, photinos remain in thermal equilibrium
much longer than in conventional SUSY, due to pion catalysis of their
conversion to $R^0$'s: $\pho \pi \leftrightarrow \r0 \pi$.  The
$\r0$'s stay in thermal equilibrium still longer, because their
self-annihilation to pions has a strong interaction cross section.
The relic abundance of photinos depends sensitively on the ratio of
the $R^0$ and $\pho$ masses\cite{f:100}.
This is because the Boltzman probability of finding a pion with
sufficient energy to produce an $\r0$ from a $\pho$ decreases
exponentially as the $\r0$ mass increases.  This was studied in sudden
approximation in ref. \cite{f:100} using the most relevant
reactions.  The analysis has been refined in ref. \cite{f:113} by
integrating the coupled system of Boltzman equations for the reactions
$\pho \pi \leftrightarrow \r0 \pi$, $\r0 \leftrightarrow \pi^+ \pi^-
\pho$, $\r0 \pho \leftrightarrow \pi^+ \pi^- $, and $\r0 \r0$ total
annihilation.  Defining $r \equiv \frac{M(R^0)}{m_\pho}$, ref. \cite{f:113}
finds that for relic photinos to give $\Omega h^2 \sim 0.25$ requires
$1.3 \la r \la 1.55$.  Assuming photinos are stable, their relic mass density
would be unacceptably large unless $r \la 1.8$.  These values of $r$ are consistent
with the mass estimates quoted above for the $\r0$ and $\pho$, which encourages 
us to take the possibility of light photinos seriously.

The detectability of relic dark matter is different for light $\pho$'s
than in the conventional heavy WIMP scenario for two reasons.
The usual relation between the relic density and the
WIMP-matter scattering cross section only applies when the relic
density is determined by the WIMP self-annihilation cross section.  In
order to have the correct relic abundance, the rate of photino-removal
from the thermal plasma must be greater than the expansion rate of the
Universe until a temperature of order $m_\pho/22$.  In the light photino
scenario, the photino relic density is determined by the $\r0 - \pho$
interconversion cross section.  When the dominant process keeping
photinos in thermal equilibrium is interconversion ($\Gamma =
n_\pi <\sigma_{R \pi \leftrightarrow \pho \pi} v>$) the required 
cross section, now $\sigma_{R \pi \leftrightarrow \pho
\pi}$, is smaller than when equilibrium is maintained by photino
annihilation ($\Gamma = n_\pho <\sigma_{\pho \pho \leftrightarrow f \bar{f}}
v>$), because $n_\pi >> n_\pho$\cite{f:100}.  Furthermore, $\sigma_{R \pi
\leftrightarrow \pho \pi}$ is parametrically larger than $\sigma_{\pho \pho
\leftrightarrow f \bar{f}}$ by the factor $\alpha_s/\alpha$ and does not
require a $p$-wave initial state.  The photino-matter scattering
cross section is therefore correspondingly smaller than in the usual WIMP
scenario.  Goodman and Witten in Ref. \cite{goodwitt} discuss $\pho$ detection
through $\pho$-nucleon elastic scattering. Using Eq.\ (3) of Ref. \cite{goodwitt}
and the parameters for light photinos, one finds event rates between
$10^{-3}$ and $10$ events/(kg day)\cite{f:113}.    

Even if the event rate were larger, observation of relic light
photinos would be difficult with existing detectors because the
sensitivity of a generic detector is poor for the $\la 1$ GeV photino
mass relevant in this case, because WIMP detectors have generally been 
optimized to maximize the recoil energy for a WIMP mass of order $10$
to $100$ GeV.   

\begin{figure}
\hspace*{25pt} \epsfxsize=400pt \epsfbox{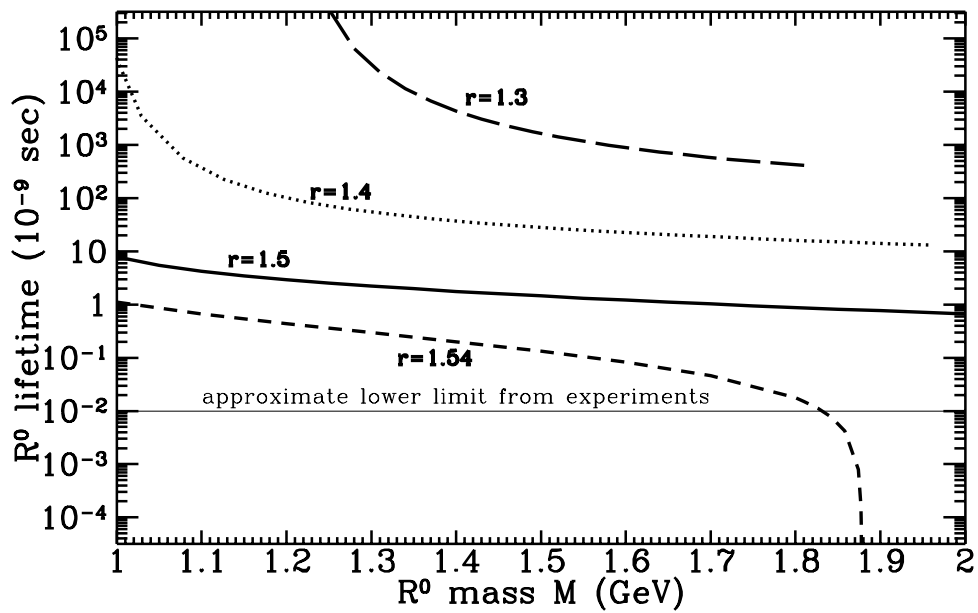}\\
\caption{$R^0$ lifetime (strictly speaking, lower limit thereto) as a
function of the $R^0$ mass, for several values of $r \equiv
M_\r0/m_\pho$, when $\Omega_\pho h^2 = 0.25$. }
\label{fig:tau}
\end{figure}

The amplitudes for the reactions responsible for $R^0$ decay
($\r0 \leftrightarrow \pi^+ \pi^- \pho$) and for the photino 
relic density ($\pho \pi \leftrightarrow \r0 \pi$) are related by
crossing symmetry.  If the momentum dependence of the amplitude is 
mild, the $\r0$ lifetime and the photino relic abundance depend
on a single common parameter in addition to the $\r0$ and $\pho$
masses\cite{f:113}.  In that case, demanding the correct
dark matter density determines the $\r0$ lifetime given the 
$\r0$ and $\pho$ masses.  The resulting lifetimes are shown in Fig.
\ref{fig:tau} from \cite{f:113}.  In actuality, the interconversion
reaction $\pho \pi \leftrightarrow \r0 \pi$ is expected to have a
resonance, so momentum-independence of the amplitudes is not a good
assumption for all of parameter space.  However this merely lengthens
the $\r0$ lifetime in comparison with the crossing relation Fig.
\ref{fig:tau}.  The required lifetime range is consistent with both
the experimental limits\cite{f:95,f:119} and with the predicted range of
lifetimes, so relic light photinos pass an important hurdle.

\begin{figure}
\vspace{9pt}
\epsfxsize = 400pt \epsffile{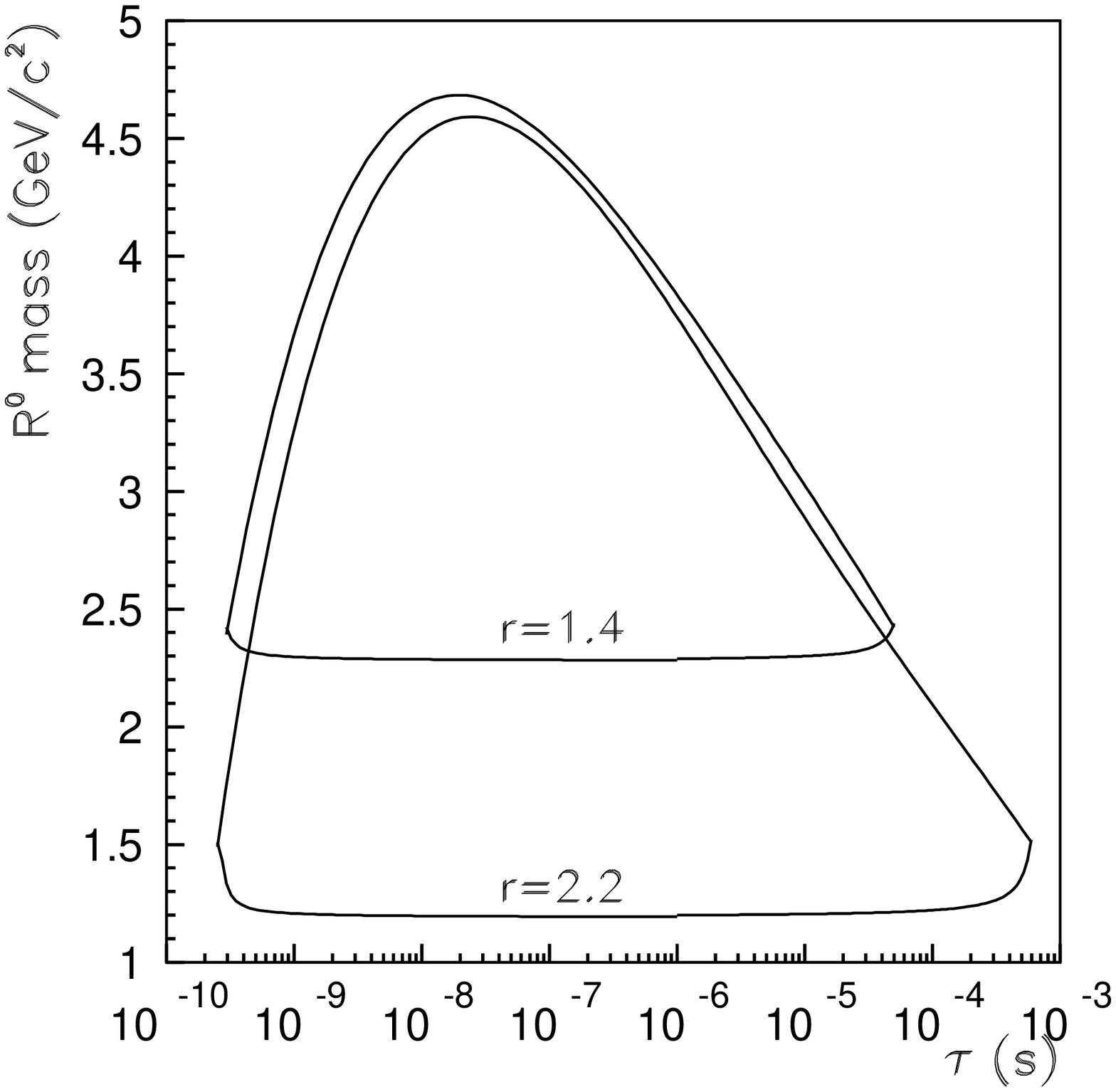}
\caption[]{KTeV limits:  inside the triangular regions is excluded\cite{ktev:lg}.} 
\label{ktev}
\end{figure}

Using data from 1 day of running, KTeV\cite{ktev:lg} has placed limits on
the production of $R^0$'s and their subsequent decay via $R^0 \rightarrow
\pi^+ \pi^- \pho$, as proposed in \cite{f:104}.  
Their cut $m(\pi^+ \pi^-) > 648$ MeV (designed to eliminate background
from $K_L^0$ decays) restricts them to the study of the kinematic region
$m(R^0)(1-1/r) > 648$ MeV.  Subject to this constraint their limits
are extremely good.  Fig. \ref{ktev} shows the mass-lifetime
region excluded by KTeV, for two values of $r$. For the largest
$r$ allowed by cosmology, 1.8, KTeV is sensitive to $m(R^0)\gsi 1\frac{1}{2}$
GeV and considerably improves the previous limits\cite{f:95}.  However 
the sensitivity drops rapidly for lower $r$ and the analysis is completely blind
to the $R^0$ mass region of primary interest for $r \le 1.4$.
The experimental challenge will be to reduce the invariant mass cut, in order to
obtain limits which are relevant to the photino-as-dark-matter question,
for which $1.3 \lsi r \lsi 1.55$.

In order to naturally account for dark matter with light photinos as
discussed in this section, all gaugino masses must be small.  This means
that some of the charginos and neutralinos (the eigenstates of
the higgsino-gaugino mass matrix) must be lighter than the $W$ and
$Z$, so they can be pair-produced at LEP2 energies.  Ref. \cite{f:116} describes
how this can be studied when the presence of light gluinos modifies the chargino
and neutralino branching ratios usually assumed in SUSY searches.  Applying
this technique, OPAL\cite{opal:sighad} has already completely excluded this
scenario for the extreme case that all charginos and heavier neutralinos decay
100\% of the time to purely hadronic final states.  In the next
year or so, improved statistical power and higher energy should either
exclude or yield evidence in favor of the all-gauginos-light scenario for
the general case of arbitrary final states\cite{f:116}.

The possibility that the anomalous 4 jet events observed by ALEPH\cite{aleph:4j}
arise from pair production of a $\approx 55$ GeV chargino, with the chargino
decaying via a real or virtual squark to $q \bar{q} \gl$, has not yet been
excluded. Under this interpretation, the excess events would in actuality contain 6
jets, which produce the peak in total dijet invariant mass because of the
analysis procedure\cite{ch4j}. Due to the method of determining 
the dijet masses, such a peak would probably disappear with increasing energy,
as has apparently occured in the data.  A Monte Carlo of the 6-jet process is
needed in order to decide the question.  Of course the method described in
the preceeding paragraph, when extended to arbitrary branching fractions,
could also be used to exclude this possibility.  The branching fraction
required to account for the apparent cross section at 130 GeV, $\approx
1/2$ pb averaging over all 4 experiments, is too low to have been
excluded by the OPAL analysis\cite{opal:sighad} discussed above.

\section*{Ultra-High-Energy Cosmic Rays}
\hspace*{2em}
If the light gaugino scenario is correct, the lightest $R$-baryon,
$S^0 \equiv uds\glu$, may be responsible for the very highest energy 
cosmic rays reaching Earth.  Recall that the observation of several
events with energies $\ga 2~ 10^{20}$ eV\cite{akeno_flyseye} presents
a severe puzzle for astrophysics\footnote{For a recent survey and
references see \cite{uhecr}.}.  Protons with such high energies have a
large scattering cross section on the 2.7 K microwave background
photons, because $E_{cm}$ is sufficient to excite the $\Delta(1230)$
resonance\cite{GZK}.  Consequently the scattering length
of such high energy protons is of order 10 Mpc or 
less.  The upper bound on the energy of cosmic rays which could have
originated in the local cluster, $\sim 10^{20}$ eV, is called the
Greisen-Zatsepin-Kuzmin (GZK) bound.  

\begin{figure}[p]
\hspace*{25pt} \epsfxsize=400pt \epsfbox{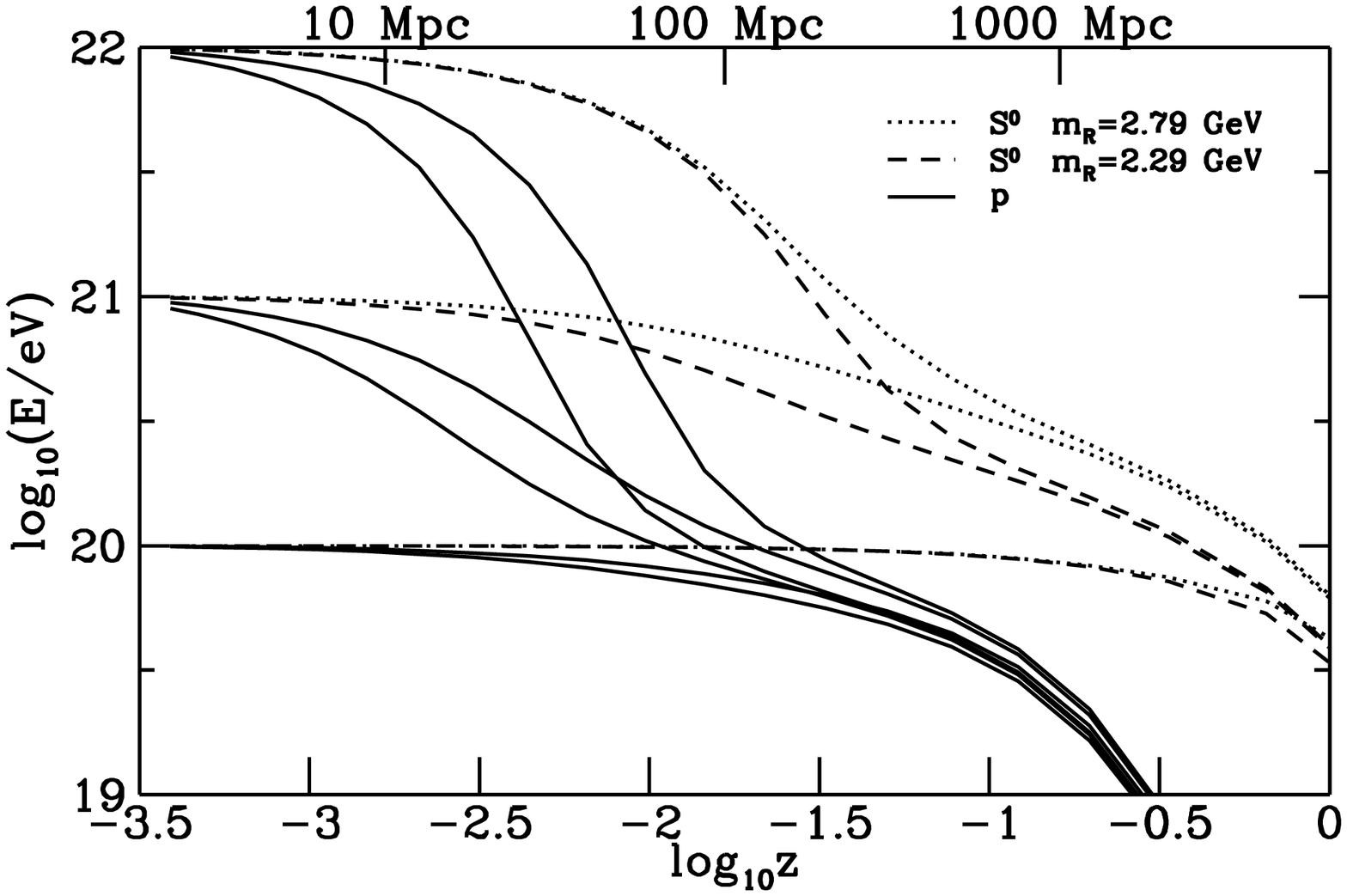}\\
\hspace*{25pt} \epsfxsize=400pt \epsfbox{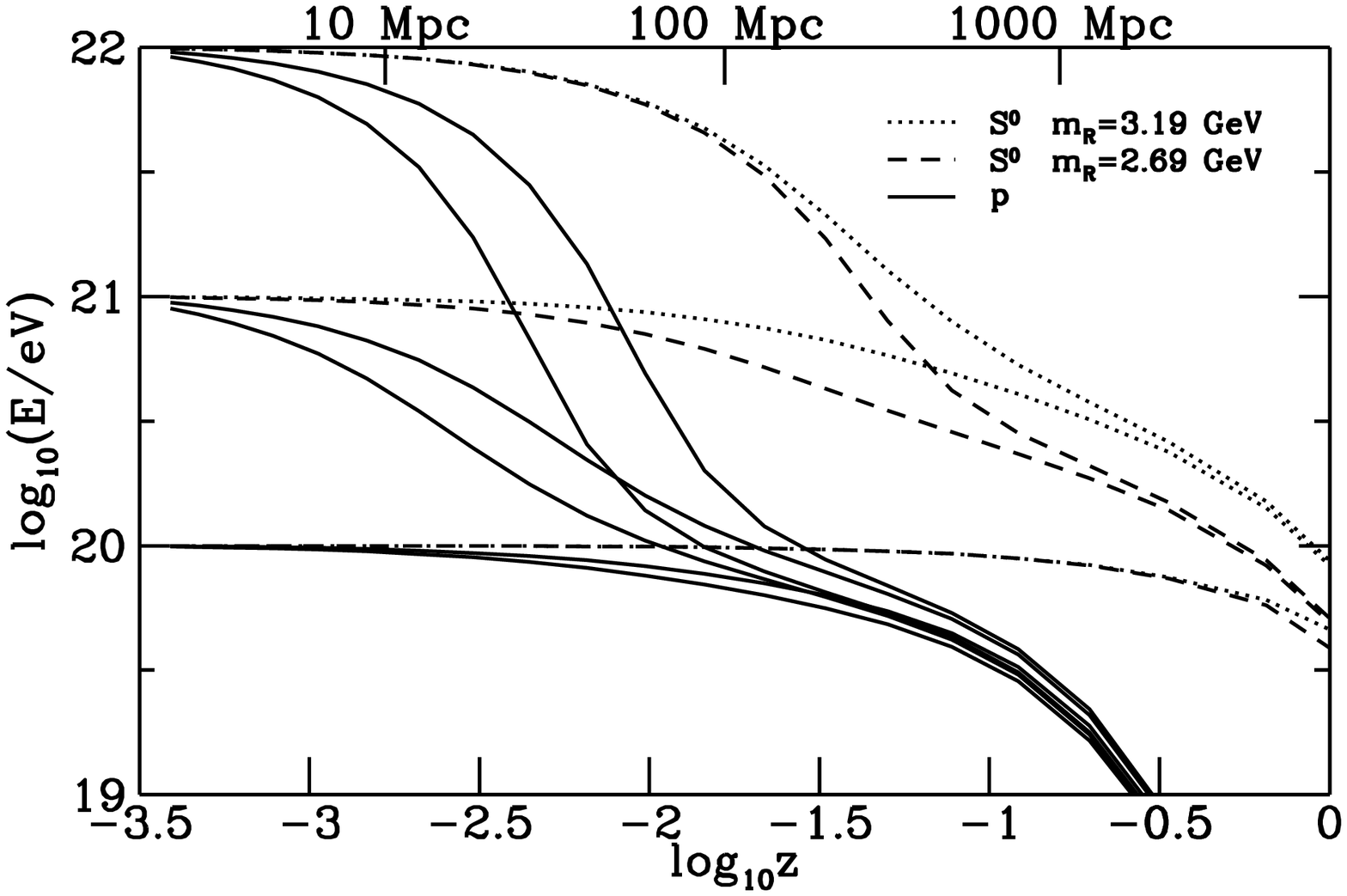}\\
\caption[]{ The figures show the
primary particle's energy as it would be observed on Earth today if it
were injected with various energies ($10^{22}eV$ eV, $10^{21}$ eV, and
$10^{20}$ eV) at various redshifts.  The distances correspond to
luminosity distances.  The mass of $S^0$ is $1.9 $GeV in the upper plot
while it is $2.3 $GeV in the lower plot.  
 Here, the Hubble constant has been set to 50
km sec$^{-1}$ Mpc$^{-1}$. }
\label{gzk}
\end{figure}

Two of the highest energy cosmic ray events come from the same
direction in the sky\cite{uhecr}; the geometrical random probability for this
is $\sim 10^{-3}$.  The nearest plausible source in that direction is the
Seyfert galaxy MCG 8-11-11 (aka UGC 03374), but it is 62-124 Mpc
away\cite{ElbSom}.  An even more plausible source is the AGN 3C 147, but
its distance is at least a Gpc\cite{ElbSom}.  The solid curves in Fig.
\ref{gzk}, reproduced from ref. \cite{f:114}, shows the spectrum of high
energy protons as a function of their initial distance, for several different
values of the energy.  Compton scattering and photoproduction, as well as
redshift effects, have been included.  It is evidently highly unlikely that
the highest energy cosmic ray events can be due to protons from MCG 8-11-11,
and even more unlikely that two high energy protons could penetrate
such distances or originate from 3C 147.

It is also unlikely that the UHECR primaries are photons.  First of all,
photons of these energies have a scattering length, 6.6 Mpc, comparable to that of
protons when account is taken of scattering from radio as well
as CMBR photons\cite{ElbSom}.  Secondly, the atmospheric showers appear
to be hadronic rather than electromagnetic.  The UHECRs observed via extensive
air shower detectors have large muon content and Ref. \cite{halzen} compared
the shower development expected from a photon primary to that observed for
the Fly's Eye event and concluded that the two are not compatible.  

However the ground-state $R$-baryon, the flavor singlet scalar $uds
\tilde{g}$ bound state denoted $S^0$, could explain these
ultra-high-energy events\cite{f:104}.  On account of the very 
strong hyperfine attraction among the quarks in the flavor-singlet
channel\cite{f:52}, the $S^0$ mass is about $210 \pm 20$ 
MeV lower than that of the lightest $R$-nucleons.  As long as $m(S^0)$ is
less than $m(p) + m(R^0)$, the $S^0$ must decay to a photino rather
than $R^0$.  It would have an extremely long lifetime since its decay
requires a flavor-changing-neutral-weak transition.  The $S^0$ could
even be stable, if $m(S^0) - m(p) - m(e^-) < m_{\tilde{\gamma}}$ and
baryon number is conserved\cite{f:104}.  

If the $S^0$ lifetime is longer than $\sim 10^5$ sec,\footnote{This
is the proper time required for a few $10^{20}$ GeV particle of mass
$\sim 2$ GeV to travel $\sim 100$ Mpc.} it is a good candidate
to be the UHECR primary\cite{f:104,f:114}. The GZK bound for the $S^0$ is
several times higher than for protons. Three effects contribute to this:
(a) The $S^0$ is neutral, so its interactions with photons cancel at leading
order and are only present due to inhomogeneities in its quark
substructure. (b) The $S^0$ is heavier than the proton.  (c) The mass splitting
between the $S^0$ and the lowest lying resonances which can be reached
in a $\gamma S^0$ collision (mass $\equiv M^*$) is larger than the 
proton-$\Delta(1230)$ splitting.  

The threshold energy for exciting the resonances in $\gamma S^0$
collisions is larger than in $\gamma p$ collisions by the  
factor\cite{f:104} $\frac{m_{S^0}}{m_p} \frac{( M^* -
M_{S^0})}{(1230-940) {\rm MeV}}$.  We can estimate $M_{S^0}$ and $M^*
- M_{S^0}$ as follows.  Taking $m(R^0) = 1.3 - 2.2$ GeV,
$m_{\tilde{\gamma}}$ must lie in the range $ \sim 0.9 - 1.7$ GeV if photinos
account for the relic dark matter.  In order that the $S^0$ be stable or
very long lived we require $m_{S^0} \la m_p + m_{\tilde{\gamma}}$.  On the
other hand, the $S^0$ is unlikely to be lighter than $m_{\gl} + m(\Lambda(1405))
\approx 1.5$ GeV, assuming the cryptoexotic baryon $udsg$ is either the
$\Lambda(1405)$ as conjectured in \cite{f:104}, or heavier.  This leads to
the favored mass range $1.5 \la m(S^0) \la  2.6$ GeV. 
Since the photon couples as a flavor octet, the resonances excited in
$S^0 \gamma$ collisions are flavor octets.  Since the $S^0$ has
spin-0, only a spin-1 $R_{\Lambda}$ or $R_{\Sigma}$ can be produced
without an angular momentum barrier.  There are two $R$-baryon flavor
octets with $J=1$, one with total quark spin 3/2 and the other with
total quark spin 1/2, like the $S^0$.  Neglecting the mixing between
these states which is small, their masses are about 385-460 and
815-890 MeV heavier than the $S^0$, respectively\cite{f:52}.  Thus one
qualitatively expects that the GZK bound is a factor of 2.7 - 7.5
higher for $S^0$'s than for $p$'s, depending on which $R$-hyperons are
strongly coupled to the $\gamma S^0$ system.  A more detailed calculation
of $S^0$ scattering on microwave photons can be found in \cite{f:114}.
The results for a typical choice of parameters are shown in Fig. \ref{gzk},
confirming the crude treatment given above following ref. \cite{f:104}.  

As the above discussion makes clear, any stable hadron with mass larger
than a few times the proton mass will avoid the problem of the GZK
bound.  However as pointed out in \cite{f:114}, there is also an upper bound
on the allowed mass.  This comes about because the fractional energy loss
per collision with atmospheric nuclei is of order $m_p/m_U$, where $m_U$
is the mass of the UHECR primary.  If the energy loss per collision is too
small, the shower development will not resemble that of a nucleon. Detailed
Monte Carlo simulation is necessary to pin down the maximum acceptable
mass\cite{afk}, but it seems unlikely to exceed tens of GeV. 
Therefore new heavy colored particles predicted in many extensions of
the standard model, whose masses are $\ga 100$ GeV, could not
be the UHECR primaries even if they were not excluded on other grounds.
It is quite non-trivial that the mass of the $S^0$ in the all-gauginos-light
scenario fortuitously falls in the rather narrow range required to explain
the UHECR's. 

The question of production/acceleration of UHECR's is a difficult one, even
if the UHECR primary could be a proton.  The mechanisms which have been
considered are reviewed in ref. \cite{biermann:rev}.  I merely note here
that most of the mechanisms proposed for protons have variants which work
for $S^0$'s.  Indirect production via decay of defects
or long-lived relics of the big bang proceeds by production of extremely
high energy quarks (or gluinos).  Since all baryons and $R$-baryons
eventually decay to protons and $S^0$'s respectively, the relative probability
that a quark or gluino fragments into an $S^0$ compared to a $p$ can be
expected to be of order $10^{-1}-10^{-2}$.  This estimate incorporates the
difficulty of forming hadrons with increasingly large numbers of constituents,
as reflected in the baryon to meson ratio in quark fragmentation which is
typically of order 1:10.  To be conservative, an additional possible
suppression is included because the typical mass of $R$-baryons is greater
than that of baryons.

Mechanisms which accelerate protons also generate high energy $S^0$'s,
via the production of $R_p$'s ($uud \gl$ bound states) in $pp$
collisions\cite{f:114}.  In fact, a problem with most acceleration mechanisms
which is overcome here is that astrophysical accelerators capable of producing
ultra high energy protons have such high densities that the protons are unlikely
to escape without colliding and losing energy.  In the scenario at hand,
this high $p p$ collision rate is actually advantageous for producing $R_p$'s.
These $R_p$'s decay to $S^0 \pi^+$ via a weak interaction, with lifetime
estimated to be $2~ 10^{-11} - 2~10^{-10}$ sec\cite{f:104}.   The $S^0$
cross section is likely to be smaller than that of a nucleon by up to a
factor of 10\cite{f:104}; if so, the $S^0$'s escape without further energy
loss. 

\begin{figure}
\vspace{9pt}
\epsfxsize = 400pt \epsffile{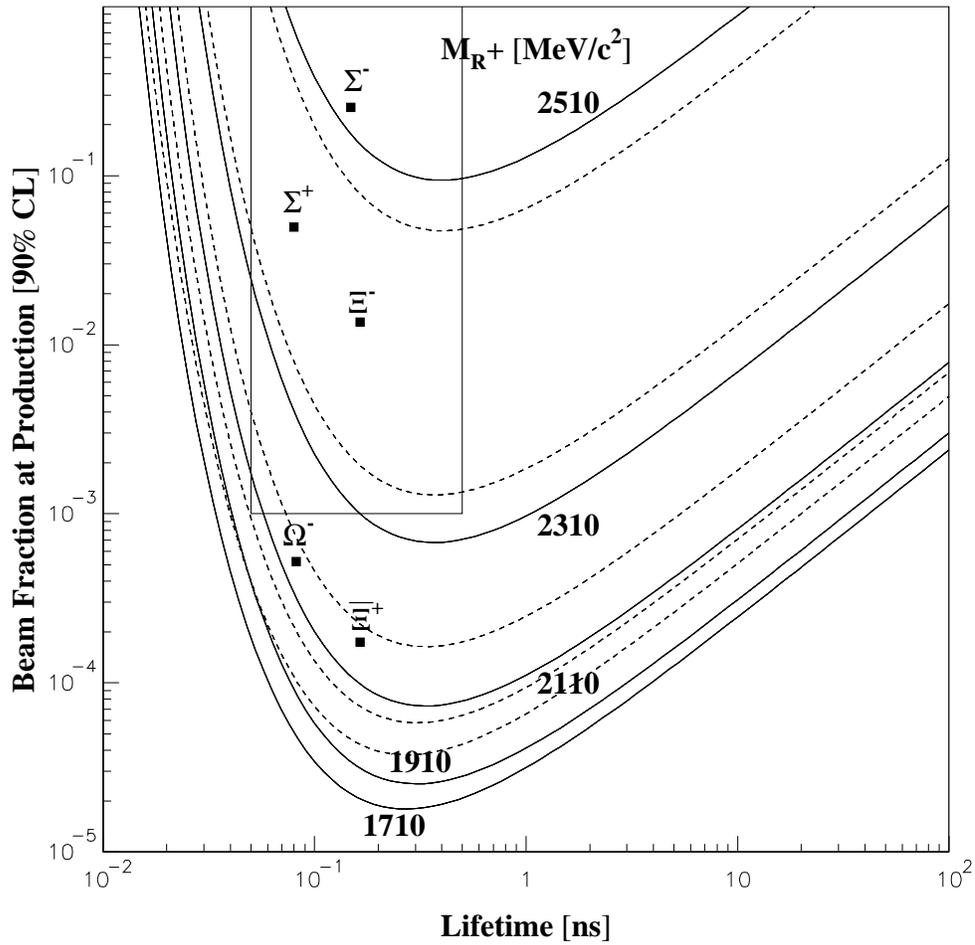}
\caption[]{E761 limits vs. $\tau(R_p)$\cite{e761}.  Ignore the box.} 
\label{e761}
\end{figure}

Laboratory experiments can be used to get upper bounds on the production
of $R_p$'s, which may be helpful in deciding whether this $S^0$ production
mechanism is plausible.  The E761 collaboration at Fermilab searched
for evidence of $R_p \rightarrow S^0 \pi^+$\cite{e761}.  Their result is
shown in Fig. \ref{e761}. If the lifetime of the $R_p$ is of order nanoseconds,
these limits would make it difficult to produce sufficient high energy $S^0$'s
via $R_p$'s. But for a lifetime of order $2~ 10^{-1} - 2~ 10^{-2}$ ns as
estimated in \cite{f:104}, the E761 limits are too weak to be a constraint.
As detailed in \cite{f:119}, a second generation experiment of this type
would be very valuable.

Finally, I note that the predicted time-dilated lifetime of a $\sim 3~ 10^{20}$
eV $R_p$, is of order seconds -- a characteristic timescale for Gamma Ray
Bursts.  Mechanisms for producing ultra high energy protons in GRB's
would translate to the production of $R_p$'s\cite{waxman:cr}.

With the much larger sample of ultra-high-energy cosmic rays expected
from HiRes and hopefully the Auger Project, the prediction of a GZK cutoff
shifted to higher energy can be tested.  The $S^0$'s are
not deflected by magnetic fields so they should accurately point
to their sources\footnote{Since the $S^0$ is a neutral spin-0 particles,
even its magnetic dipole moment vanishes.}.  That means UHECR's should cluster
about certain directions in the sky, if $S^0$'s are the primaries and the
sources are not intermittant.

\section*{Summary}

$\bullet$ Light photinos can account for the relic dark matter, if the
$R^0$ (glueballino) mass is between 1.3 and 1.55 times the photino mass.
This is consistent with the predicted mass ranges and present experimental
constraints. \\  
$\bullet$ If dark matter is due to relic photinos, one can expect
$10^{-3} - 10$ interactions per kg per day in a WIMP detector.  However since
the photino mass is of order 1 GeV in this scenario, they will not deposit 
significant energy in detectors based on heavy nuclei. \\
$\bullet$  The cosmic ray events whose energy is above the GZK bound
may be due to the lightest gluino-containing baryon, a $uds\gl$
bound state called the $S^0$.  A cutoff in the spectrum at a somewhat
higher energy is predicted, as is sharp pointing to the sources.\\
$\bullet$  Three new laboratory searches are described.  At present
their sensitivity is insufficient to test this picture, but that should
change in a year or two.




\begin{thebibliography}{10}

\bibitem{f99101}
G.~R. Farrar.
\newblock Technical Reports RU-95-17 (hep-ph/9504295) and RU-95-25
(hep-ph/9508291), Rutgers Univ., 1995.  Invited talk SUSY95, Paris, May 1995,
RU-95-73 (hep-ph/9704310). 

\bibitem{radmass}
R.~Barbieri and L.~Maiani, \newblock {\em Nucl. Phys.}, B243:429, 1984;
G.~R. Farrar and A.~Masiero,\newblock Technical Report RU-94-38,
hep-ph/9410401, Rutgers Univ., 1994; D.~Pierce and A.~Papadopoulos,
\newblock {\em Nucl. Phys.}, B430:278, 1994;
G.~R. Farrar, \newblock Technical Report RU-95-26 (hep-ph/9508292),
Rutgers Univ., 1995. 

\bibitem{f:108}
G.~R. Farrar and G.~Gabadadze.
\newblock {\em Phys. Lett.}, 397B:104, 1997.

\bibitem{f:24}
G.~R. Farrar and P.~Fayet.
\newblock {\em Phys. Lett.}, 79B:442--446, 1978.

\bibitem{f:95}
G.~R. Farrar.
\newblock {\em Phys. Rev.}, D51:3904, 1995.

\bibitem{f:104}
G.~R. Farrar.
\newblock {\em Phys. Rev. Lett.}, 76:4111, 1996.

\bibitem{f:109}
F.~E. Close, G.~R. Farrar, and Z.~P. Li.
\newblock {\em Phys. Rev.}, D55:5749, 1997.

\bibitem{f:51}
G.~R. Farrar.
\newblock {\em Phys. Rev. Lett.}, 53:1029--1033, 1984.

\bibitem{aleph:lg}
The~Aleph Collaboration.
\newblock {\em Z. Phys.}, C76:191-199, 1997.

\bibitem{fLaT} B. Gary, CTEQ Workshop, FNAL, Nov. 1996; G. R.
Farrar, Rencontres de la Valee d'Aoste, La Thuile, Feb. 1997, RU-97-22
(hep-ph/9707467).

\bibitem{f:119} G. R. Farrar, Invited Review SUSY97, Philadelphia,
May 1997, RU-97-79 (hep-ph/9710277).

\bibitem{nt:lg}  Z.~Nagy and Z.~Trocsanyi, hep-ph/9708343.

\bibitem{f:100}
G.~R. Farrar and E.~W. Kolb.
\newblock {\em Phys. Rev.}, D53:2990, 1996.

\bibitem{f:113}
D.~J. Chung, G.~R. Farrar, and E.~W. Kolb.
\newblock Technical Report FERMILAB-Pub-96-097-A, RU-97-13 and
astro-ph/9703145, FNAL and Rutgers Univ, 1997.  Phys. Rev. D. to be
published. 

\bibitem{goodwitt}
M.~W. Goodman and E.~Witten.
\newblock {\em Phys. Rev.}, D31:3059, 1985.

\bibitem{ktev:lg}
The~KTeV Collaboration.
\newblock  RU-97-26, Rutgers Univ., 1997, Phys. Rev. Lett., to
  be published.

\bibitem{f:116}
G.~R. Farrar.
\newblock  RU-97-20 (hep-ph/9706393) Rutgers Univ., 1997.

\bibitem{opal:sighad}
The~OPAL Collaboration.
\newblock  PPE 97-101, hep-ex/970802, CERN, 1997.

\bibitem{aleph:4j}
The~Aleph Collaboration.
\newblock {\em Z. Phys.}, C71:179-198, 1996.

\bibitem{ch4j}
G.~R. Farrar.
\newblock {\em Phys. Rev. Lett.}, 76:4115, 1996 and also
Technical Report invited talk Warsaw ICHEP96, RU-96-93 
(hep-ph/9612355) Rutgers Univ., 1996.

\bibitem{akeno_flyseye}
N.~Hayashida.
\newblock {\em Phys. Rev. Lett.}, 73:3491, 1994; D.~J. Bird et~al.
\newblock {\em Ap. J.}, 441:144, 1995.

\bibitem{uhecr}
J.~Lloyd-Evans and A.~Watson.
\newblock {\em Phys. World}, 9:47, 1996.

\bibitem{GZK}
K.~Greisen.
\newblock {\em Phys. Rev. Lett.}, 16:748, 1966; 
G.~T. Zatsepin and V.~A. Kuzmin.
\newblock {\em Sov. Phys.-JETP Lett.}, 4:78, 1966.

\bibitem{ElbSom}
J.~Elbert and P.~Sommers.
\newblock {\em Ap. J.}, 441:151, 1995.

\bibitem{f:114}
D.~J. Chung, G.~R. Farrar, and E.~W. Kolb.
\newblock Technical Report RU-97-14 (astro-ph/9707036) FNAL and  Rutgers
Univ, 1997. 

\bibitem{halzen}
F.~Halzen et al.
\newblock {\em Astroparticle Phys.}, 3:151, 1995.

\bibitem{f:52}
F.~Bucella, G.~R. Farrar, and A.~Pugliese.
\newblock {\em Phys. Lett.}, B153:311--314, 1985.

\bibitem{afk}
I.~F. Albuquerque, G.~R. Farrar, and E.~W. Kolb.
\newblock Work in progress.

\bibitem{biermann:rev}
P.~Biermann.
\newblock {\em J. Phys. G: Nucl. Part. Phys.}, 23:1-27, 1997.

\bibitem{e761}
I.~F. Albuquerque and others (E761~Collaboration).
\newblock {\em Phys. Rev. Lett.}, 78:3252, 1997.

\bibitem{waxman:cr}
\newblock Proc. ICRR Symposium on Extremely High Energy Cosmic Rays, ed.
M. Nagano, Tokyo, 1996.  Technical report IASSNS-AST 96/61.

\end{thebibliography}

\end{document}